\documentclass[12]{article}
\usepackage[dvips]{graphicx}

\def\bx{{\bf x}}
\def\bu{{\bf u}}
\def\bv{{\bf v}}
\def\by{{\bf y}}

\def\ba{{\bf a}}
\def\bb{{\bf b}}
\def\bc{{\bf c}}
\def\be{{\bf e}}
\def\bd{{\bf d}}

\def\bv{{\bf v}}

\def\by{{\bf y}}
\def\bw{{\bf w}}

\def\wt{\widetilde}
\def\C{{C\kern-.647em I}}
\def\H{I\!\!H}

\def\R{I\!\!R}

\def\beq{\begin{equation}}
\def\eeq{\end{equation}}

\def\w{\wedge}

\def\C{{I\! \! \! C}}

\def\G{I\!\!\!G}

\def\wt{\widetilde}

\def\beq{\begin{equation}}
\def\eeq{\end{equation}}

\def\G{I\!\!\!G}

\def\wt{\widetilde}

\def\ba{{\bf a}}
\def\bb{{\bf b}}
\def\bc{{\bf c}}
\def\bd{{\bf d}}

\def\bv{{\bf v}}

\def\bx{{\bf x}}
\def\by{{\bf y}}
\def\ba{{\bf a}}
\def\bb{{\bf b}}

\def\bc{{\bf c}}

\def\R{I\!\!R}
\def\H{I\!\!H}
\def\C{I\!\!\!C}

\begin{document}

\title{Geometry of Moving Planes}

\author{Garret Sobczyk, \\
Departamento de Actuar\'ia y Matem\'aticas \\ Universidad de Las Am\'ericas - Puebla,\\ 72820 Cholula, Mexico}

\maketitle

\begin{abstract}
   The concept of number and its generalization has played a central role in the development of
mathematics over many centuries and many civilizations. Noteworthy milestones in this long and
arduous process were the developments of the real and complex numbers which have achieved
universal acceptance. Serious attempts have been made at further extensions,
such as Hamiltons quaternions, Grassmann's exterior algebra and Clifford's geometric algebra. 
By examining the geometry of moving planes, we show how new mathematics is within reach,
if the will to learn these powerful methods can be found. 

 \end{abstract}
%\centerline{\bf \Large OUTLINE}

\section*{Introduction}

    Great advances in mathematics have been made by repeated extensions of the
    concept of number. 
   The real number system $\R$ has a long and august history spanning a host of civilizations over
many centuries. It is the rock upon which many other mathematical systems are
constructed, and serves as a model of desirable properties that a number system
 should have. A property which the real number system does not have, the {\it closure property}
for the zeros of any real polynomial, historically provided one the most compelling reasons
for their extension. By extending the real numbers $\R$ to
include an {\it imaginary unit} $i:= \sqrt{ -1} \notin \R$, we arrive at the {\it complex numbers} $\C$.
The complex numbers enjoy all the algebraic properties of the reals, but in addition are
algebraically closed. Any complex number $z\in \C$ can be expressed 
in the {\it standard basis} $\{1, i\}_{\R}$ as $z=x+i y$ where
$x,y\in \R$, leading to the idea of the $2$-dimensional {\it complex number plane}.

   Over the last 150 years a rich {\it complex analysis} has been developed, which has been 
fully incorporated into the mathematical toolbox of every mathematician and practitioners
of mathematics from the engineering and scientific communities. The famous {\it Euler formula} 
  \beq \exp(i \theta) = \cos \theta + i \sin \theta \label{eulertrig} \eeq
helps mades clear the geometric significance of the
multiplication of complex numbers. 

   Whereas the complex numbers have enjoyed
universal acceptance and admiration, other extensions have met with
greater resistance and have found only limited acceptance. For example, the extension
of the complex numbers to Hamilton's quaternions, has been more divisive in its effects upon the
mathematical community \cite{Crowe}. Other powerful extensions, such as Grassmann's exterior algebras
and William K. Clifford's geometric algebras \cite{Clifford}, have had a profound effect on the development
of higher mathematics, but have yet to be brought into the mainstream of mathematics. A
revealing history is told in \cite[pp.320-327]{LP01}, and a 
website devoted to telling this fascinating story, with many references to the literature,
 can be found at (http://modelingnts.la.asu.edu/).

   The principal roadblock to further extensions of the real number system has been the failure to
consider the extension of the real numbers to include {\it new} square roots of $+1$, perhaps 
because such considerations were for the most part 
   before the advent of Einstein's {\it theory of special relativity} and the study of {\it non-Euclidean}
   geometries. Extending the real number system $\R$ to include a new
   square root $u=\sqrt 1 \notin \R$ leads to the concept of the {\it hyperbolic number plane} $\H$, which in many ways
   is analogous to the complex number plane $\C$. Understanding the hyperbolic numbers 
   is key to understanding even more general {\it geometric extensions} of the real numbers.   

   A {\it hyperbolic number} $w \in \H$, in the {\it standard basis} $\{1, u\}_{\R}$, has the form $w=x+u y$ for $x,y \in \R$.
The hyperbolic numbers $\H$ enjoy all the properties of the real numbers $\R$, except that $\H$ has zero divisors.
The real hyperbolic numbers $\H$ have the structure of a commutative ring, but are not algebraically closed.
It is interesting to note that the hyperbolic numbers, just like the complex numbers, can be used to derive
the not-so-well known formula for the zeros of a real cubic polynomial \cite{S1}.

   The {\it Euler forms} of a hyperbolic number $w=x+uy \in \H$ are $w=\pm \rho \exp{u \phi}$ or 
$w=\pm \rho u \exp{u \phi}$ for $\rho=\sqrt{|x^2-y^2|}$ and $\phi=Tanh^{-1}(y/x)$ or 
$\phi=Tanh^{-1}(x/y)$, respectively, corresponding to the $4$ branches of the
unit hyperbola $x^2-y^2=\pm 1$. Expanding $e^{u\phi}$ in terms of the hyperbolic trig functions gives
  \beq  e^{u \phi}=\cosh \phi + u \sinh \phi, \label{eulerhyp}  \eeq
which of course is analogous to (\ref{eulertrig}).  
The Euler forms facilitate the geometric interpretation of the multiplication
of hyperbolic numbers. For example, if $w_1 = \rho_1 e^{u \phi_1}$ and $w_2 = \rho_2 e^{u \phi_2}$, then
  \[ w_1 w_2 = \rho_1 \rho_2 e^{u(\phi_1+\phi_2)} . \]
The {\it hyperbolic distance} between $w_1, w_2\in \H$ is defined by 
\[|w_1-w_2|=\sqrt{|(x_1-x_2)^2-(y_1-y_2)^2|}, \]
and the equation of the hyperbola with hyperbolic
radius $\rho$ is $|w w^-| =|x^2-y^2|=\rho^2$, where $w^- :=x-yu$.
 
   The defect that the hyperbolic numbers are not algebraically closed can be
  remedied by introducing the $4$-dimensional {\it complex hyperbolic numbers}. But instead,
following William Kingdon Clifford \cite{Clifford},  
  we will consider the extension of the real numbers obtained by introducing two
  {\it anticommuting} square roots of $+1$. 

  \section*{Geometric numbers of the $2$-plane}

To obtain the associative geometric algebra $\G_2$ of the Euclidean plane $\R^2$, we extend the real numbers $\R$ to include
two new {\it anticommuting} square roots $\be_1$ and $ \be_2$ of $+1$, so that
\[ \G_2=span \{1, \be_1, \be_2, \be_{12} \}_{\R},  \] 
where $\be_1^2=\be_2^2=1$ and $\be_{12}:=\be_1 \be_2=-\be_2 \be_1$. We say that 
  \beq {\cal B}:= \{1, \be_1, \be_2, \be_1 \be_2 \}  \label{stdbasis} \eeq
is a {\it standard orthonormal basis} of $\G_2$, where $\be_1$ and $\be_2$ are given the
interpretation of
{\it orthonormal vectors} along the $x$ and $y$ axis of $\R^2$, respectively.
The quantity $i:=\be_{12}$ has the geometric interpretation of a {\it unit
bivector}, and defines the {\it direction} and {\it orientation} of the vector plane $\R^2$. 
See Figure 1. The geometric algebra  $\G_2$ obeys all the algebraic rules of the real numbers $\R$, except 
that $\G_2$ has zero divisors and is not universally commutative.

\begin{figure}
\begin{center}
\includegraphics[scale=.50]{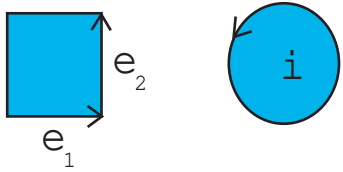}
\caption{The unit bivector $i$ of $\R^2$.}
\end{center}
\end{figure} 

Calculating 
  \[ i^2=(\be_1 \be_2)(\be_1 \be_2)=-\be_1 (\be_1 \be_2) \be_2 = -\be_1^2 \be_2^2= -1, \]
we see that $i$ has the same algebraic property as the imaginary unit of the complex numbers. 
    The most general geometric number $g\in \G_2$ has the form
   \beq g=(x + i y) + (v_1 \be_1+v_2 \be_2)=z+\bv, \label{gform}  \eeq
where the {\it spinor} 
$z=x+iy$ for $x,y \in \R$ behaves like a complex number, and $\bv=v_1\be_1+v_2 \be_2 \in \R^2$
is a {\it vector} in the two dimensional Euclidean space $\R^2$.
We say that $z\in {\cal S}_i$ where ${\cal S}_i$ is the {\it spinor plane} of the bivector $i=\be_1 \be_2$.
Noting that 
\beq \be_1 i = \be_1 \be_1 \be_2 = \be_2,\ {\rm and} \ \be_2 i = \be_2 \be_1 \be_2 = -\be_1  \label{90degrees} \eeq
it follows that multiplying any vector $\bv\in \R^2$ on the right by $i$ rotates the vector $\pi/2$ radians {\it counterclockwise}
in the plane of bivector $i$. Consequently, the {\it spinors} $e^{i\theta}$ 
generate rotations in the oriented vector plane of $\R^2$.

Let $\ba, \bb \in \R^2$ be two unit vectors so that $\ba^2=\bb^2=1$. Then
   \beq \ba \bb = \ba \cdot \bb + \ba \w \bb =\cos \theta + i \sin \theta =e^{i\theta},
                           \label{geoproduct}          \eeq
where $\ba \cdot \bb :=\frac{1}{2}(\ba \bb + \bb \ba)=\cos \theta$ is the
{\it symmetric inner product} and
 $\ba \w \bb :=\frac{1}{2}(\ba \bb - \bb \ba)=i\sin \theta$ is the
{\it anti-symmetric outer product} of the vectors $\ba$ and $\bb$. The equation
(\ref{geoproduct}) shows the deep relationship that exists between the vectors of
$\R^2$ and the spinors ${\cal S}_i$ of $\R^2$. We write $\R^2=\R_i^2$ to emphasize
that $\R^2$ consists of all vectors in the plane of the bivector $i=\be_1 \be_2$. 

Let $\bx \in \R^2$. Since
   \[ \bb=(\bb \ba)\ba =(\bb \ba)^{\frac{1}{2}}\ba (\ba \bb)^{\frac{1}{2}} =e^{-\frac{i\theta}{2}}\ba 
e^{\frac{i\theta}{2}},   \]
it follows that 
\beq \bx^\prime = R(\bx)= (\bb \ba)^{\frac{1}{2}}\bx (\ba \bb)^{\frac{1}{2}}=e^{-\frac{i\theta}{2}}\bx 
e^{\frac{i\theta}{2}}  \label{rotation} \eeq
defines a rotation of the vector $\bx$ in the plane of the bivector $i=\be_1 \be_2$
 into the vector $\bx^\prime$, and in particular $R(\ba)=\bb$. When applied to the
bivector $i=\be_1 \be_2$ we find that $R(i)=i$, so that a rotation of the bivector $i$ in the plane of
$i$ leaves the bivector unchanged, as expected. Note that
the {\it half-angle} version on the right hand side of (\ref{rotation}) is useful because it extends
immediately to rotations in $\R^n$. See \cite{H/S} and \cite{LP01} %\cite{P95}
 for an extensive
treatment of geometric algebras in higher dimensional pseudo-Euclidean spaces. 

  The geometric algebra $\G_2$ algebraically unites the spinor plane ${\cal S}_i$ and the vector plane $\R_i^2$ and
opens up many new possibilities. Consider the transformation $L(\bx)$ defined by
  \beq  \bx^\prime=L(\bx)=e^{-\frac{\phi \ba}{2}}\bx e^{\frac{\phi \ba}{2}} \label{boost} \eeq
for a unit vector $\ba\in \R^2$. The transformation (\ref{boost}) has
 the same half-angle form as the rotation (\ref{rotation}). 
 We say that (\ref{boost})
defines an {\it active Lorentz boost} of the vector $\bx$ into the {\it relative vector} $\bx^\prime$  moving with
velocity 
  \[ \frac{\bv}{c}=\tanh{(\phi \ba)}=\ba \tanh \phi \]
where $c$ is the {\it velocity of light}. For simplicity
we shall always take $c=1$. An active rotation and an active boost are pictured in Figure 2.
\begin{figure}
\begin{center}
\includegraphics[scale=.40]{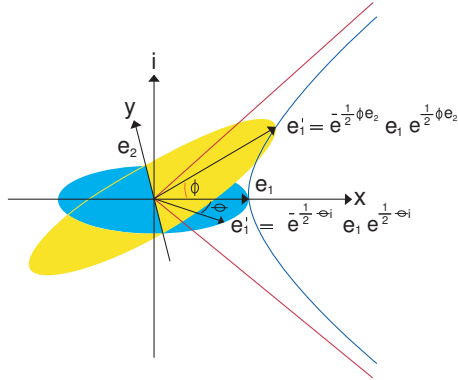}
\caption{An active rotation and an active boost.}
\end{center}
\end{figure} 

Both $R(\bx)$ and $L(\bx)$ are {\it algebra inner automorphisms} on $\G_2$,
satisfying $R(g_1 g_2)=R(g_1)R(g_2)$ and $L(g_1 g_2)=L(g_1)L(g_2)$ for all $g_1, g_2 \in \R^2$.
In addition, we say that $R(\bx)$ is an {\it outermorphism} because it preserves the {\it grading}
of the algebra $\G_2$, {\it i.e.}, for $\bx \in \R^2$, $R(\bx)\in \R^2$. Whereas a 
boost is an automorphism, it is not an outermorphism as we shall shortly see.

  Note that under both a Euclidean rotation (\ref{rotation}) and under an active boost (\ref{boost}),
  \[ |\bx^\prime|^2:=(\bx^\prime)^2=\bx^2:=|\bx|^2,   \]
so that the {\it Euclidean lengths} $|\bx|=|\bx^\prime|$ of both the rotated vector
and the boosted relative vector are preserved . Whereas the meaning
of this statement is well-known for rotations, the corresponding statement for
a boost needs further explanation.

   The active boost (\ref{boost}) leaves invariant the direction of the boost, that is
\beq L(\ba)=e^{-\frac{\phi \ba}{2}}\ba e^{\frac{\phi \ba}{2}} = \ba .  \label{invardir} \eeq
On the other hand, for the vector $\ba i$ orthogonal to $\ba$, we have 
\beq L(\ba i )=e^{-\frac{\phi \ba}{2}} \ba ie^{\frac{\phi \ba}{2}} =\ba  ie^{\phi \ba} 
      = \ba i \cosh\phi - i \sinh \phi ,  \label{orthodir} \eeq
showing that the boosted relative vector $L(\ba i)$ has picked up the bivector component
$-i \sinh \phi$. 

We say that two relative vectors are {\it orthogonal} if they are
anticommutative. From the calculation
  \beq  L(\ba)L(\ba i )=\ba \ba i e^{\phi \ba} = -\ba i e^{\phi \ba} \ba =  - L(\ba i ) L(\ba) , \label{relorthog} \eeq
  we see that the active boost of a pair orthornormal vectors gives a pair of orthornormal relative
  vectors.  When the active Lorentz boost is applied to the
bivector $i=\be_1 \be_2$ we find that $j=L(i)=ie^{\ba \phi}$, so that a boost of the bivector $i$ in the direction
of the vector $\ba$ gives the {\it relative bivector} $j= ie^{\ba \phi}$. Note that 
  \[ j^2=ie^{\ba \phi}ie^{\ba \phi}=i^2 e^{-\ba \phi}e^{\ba \phi}=-1  \]
  as expected.

       Using equations (\ref{invardir}),  (\ref{orthodir}) and (\ref{relorthog}), we say
  that 
  \[ {\cal B}_j := \{1, \be_1^\prime, \be_2^\prime, \be_1^\prime \be_2^\prime \},  \]
where $\be_1^\prime = \ba$, $\be_2^\prime= \ba i e^{\phi \ba}$, 
and $j= \ba  \ba i e^{\phi \ba}=  i e^{\phi \ba}$, 
makes up a {\it relative orthonormal basis} of $\G_2$. Note that the defining rules for the standard basis
(\ref{stdbasis}) of $\G_2$ remain the same for the relative basis ${\cal B}_j$:
  \[ (\be_1^\prime)^2=(\be_2^\prime)^2=1, \ {\rm and}\ \be_1^\prime \be_2^\prime =-\be_2^\prime \be_1^\prime. \]
 Essentially, the
 relative basis ${\cal B}_j$ of $\G_2$ {\it regrades} the algebra into relative vectors and relative
 bivectors moving at the velocity of $\bv=\ba \tanh \phi$ with respect to the standard basis ${\cal B}_i$. 
We say that $j$ defines the {\it direction} and {\it orientation} of the relative plane 
 \beq {\R_j^2} :=\{ \bv^\prime|\ \ \bv^\prime =x^\prime \be_1^\prime + y^\prime \be_2^\prime, 
   {\rm for} \ \ 
 x^\prime,  y^\prime \in \R \} . \label{relplane}  \eeq 
 
   Active rotations (\ref{rotation}) and active boosts (\ref{boost}) define two different kinds of automorphisms on the
   geometric algebra $\G_2$. Whereas active rotations are well understood in Euclidean geometry, 
   an active boost brings in concepts from non-Euclidean geometry. Since an active boost
   essentially regrades the geometric algebra $G_2$ into relative vectors and relative bivectors, it
   is natural to refer to the {\it relative geometric algebra} $\G_2$ of the relative plane (\ref{relplane})
   when using this basis.
   
   \section*{Relative geometric algebras}
  
  We have seen that both the unit bivector $i$ and the relative unit bivector $j=ie^{\ba \phi}$ 
have square $-1$. Let us see what can be said about the most general element $h \in \G_2$ which has the 
property that $h^2=-1$. In the standard basis (\ref{stdbasis}), $h$ will have the form
   \[ h=h_1 \be_1+h_2 \be_2+ h_3 i \]
for $h_1,h_2,h_3 \in \R$ as is easily verified. Clearly the condition that $h^2=h_1^2+h_2^2-h_3^2=-1$ will be satisfied
if and only if $1+h_1^2+h_2^2=h_3^2$ or $h_3=\pm \sqrt{1+h_1^2+h_2^2}$. We have two cases:
\begin{itemize}
\item[1.] If $h_3 \ge 0$, define $\cosh \phi = \sqrt{1+h_1^2+h_2^2}$, 
$\sinh \phi =\sqrt{h_1^2+h_2^2}$ and the unit vector $\ba$ such that
$i\ba \sinh \phi = h_1 \be_1+h_2 \be_2$, or $\ba=\frac{h_1 \be_2 - h_2 \be_1}{\sqrt{1+h_1^2+h_2^2}}$.
Defined in this way, $h=i e^{\ba \phi}$ is a relative bivector to $i$.
\item[2.]  If $h_3 < 0$, define $\cosh \phi = \sqrt{1+h_1^2+h_2^2}$, 
$\sinh \phi =-\sqrt{h_1^2+h_2^2}$ and the unit vector $\ba$ such that
$i\ba \sinh \phi = -(h_1 \be_1+h_2 \be_2)$, or $\ba=\frac{h_1 \be_2 - h_2 \be_1}{\sqrt{1+h_1^2+h_2^2}}$. In
this case, $h=-i e^{\ba \phi}$ is a relative bivector to $-i$.
\end{itemize}
 
   From the above remarks we see that any geometric number $h\in G_2$ with the property that $h^2=-1$ is
 a relative bivector to $\pm i$. The set of relative bivectors to $+i$,   
   \beq {\cal H}^+:= \{ i e^{\phi \ba}| \ \  \ba = \be_1 \cos \theta + \be_2 \sin \theta, \ 0 \le 
                               \theta < 2 \pi, \  \phi \in \R\} \label{posoriented} \eeq   
are said to be {\it positively oriented }. Moving relative bivectors $i$, $j$ and $k$ are pictured
in Figure 3. Similarly, the set ${\cal H}^-$ of negatively oriented
relative bivectors to $-i$ can be defined. 
\begin{figure}
\begin{center}
\includegraphics[scale=.4]{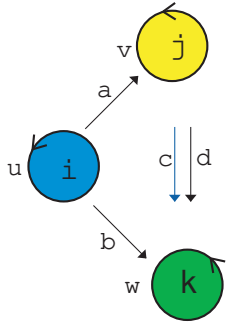}
\caption{The moving relative bivectors $i$, $j$ and $k$.}
\end{center}
\end{figure} 

   For each positively oriented relative bivector $h=i e^{\ba \phi} \in {\cal H}^+$, we define a positively
oriented relative plane $\R_h^2$ by
   \[  \R_h^2 = \{ \bx| \ \ \bx = x \ba + y \ba i, \ x,y \in \R \}, \]
and the corresponding relative basis ${\cal B}_h$ of the geometric algebra $\G_2$:
    \[ {\cal B}_h = \{1, \ba, \ba i e^{\ba \phi}, i  e^{\ba \phi} \}. \] 
In Figure 3, we have also introduced the symbols $u$, $v$ and $w$ to label the {\it systems} or
{\it oriented frames} defined by the relative bivectors $i$, $j$ and $k$, respectively. These symbols will
later take on an algebraic interpretation as well.   
    
    For each relative plane  $\R_h^2$ there exist a relative inner product and a relative outer product,
  just as in (\ref{geoproduct}). Rather than use the relative inner and outer products on
  each different relative plane, we prefer to decompose the geometric product
  of two elements $g_1, g_2 \in G_2$ into {\it symmetric} and {\it anti-symmetric} 
  parts. Thus, 
     \beq  g_1 g_2 =\frac{1}{2}(g_1 g_2+g_2 g_1)+\frac{1}{2}(g_1 g_2+g_2 g_1) 
     = g_1\circ g_2+g_1 \otimes g_2 
                                     \label{symanti} \eeq
  where $g_1 \circ g_2=   \frac{1}{2}(g_1 g_2+g_2 g_1)$ is called the {\it symmetric product} and  
  $g_1 \otimes g_2=   \frac{1}{2}(g_1 g_2-g_2 g_1)$ is called the {\it anti-symmetric product}.
  
  We give here formulas for evaluating the symmetric and anti-symmetric products of geometric
  numbers with vanishing scalar parts. Letting $A=a_1 \be_1+a_2 \be_2+a_3 i$, $B=b_1 \be_1+b_2 \be_2+b_3 i$,
  and $C= c_1 \be_1+c_2 \be_2+c_3 i$, we
  have 
     \[ A\circ B = a_1 b_1 + a_2 b_2 - a_3 b_3  \] % \label{symproduct}                                 
       \[ A\otimes B = -\det{\pmatrix{\be_1 &  \be_2 &- i \cr a_1 & a_2 & a_3 \cr b_1 & b_2 & b_3}}
             \] %\label{antisymproduct} \eeq
        \[ A\circ (B \otimes C) =- \det{\pmatrix{a_1 &  a_2 & a_3 \cr b_1 & b_2 & b_3 \cr c_1 & c_2 & c_3}},
          \] %     \label{tripleproduct} \eeq
 which bear striking resemblance to the dot and cross products of vector analysis.   
    In general, a nonzero geometric number $A\in \G_2$ with vanishing scalar part is said to be
a {\it relative vector} if $A^2>0$, a {\it nilpotent} if $A^2=0$, and a {\it relative bivector} if
$A^2<0$.  
        
 \section*{Geometry of moving planes}
 
   Consider the set ${\cal H}^+$ of positively oriented relative bivectors to $i$.
For $j\in  {\cal H}^+$, this means that $j=ie^{\phi \ba}$ as given in (\ref{posoriented}). 
We say that the system $v$, and its relative plane $\R_j^2$ defined by the bivector $j$, is 
 moving with velocity $\bu_v:=\ba \tanh \phi$ with respect to the system $u$ and its relative plane
$\R_i^2$ defined by the bivector $i$.

Note that $j=i e^{\phi \ba}$ implies that $i = j e^{-\phi \ba}$, so that if $j$ is moving
with velocity $\bu_v=\ba \tanh \phi$ with respect to $i$, then $i$ is moving
with velocity $\bv_u=-\ba \tanh \phi$ with respect to $j$. Suppose now for the system
$w$ that $k=i e^{\rho \bb} \in {\cal H}^+$,
where the unit vector $\bb \in \R_i^2$ and the hyperbolic angle $\rho \in \R$. Then
  \[ k = i (e^{\phi \ba}e^{-\phi \ba})e^{\rho \bb}=j(e^{-\phi \ba}e^{\rho \bb})=j e^{\omega \bc},  \]
where $ e^{\omega \bc}=e^{-\phi \ba}e^{\rho \bb}$ for some hyperbolic angle $\omega$ and relative unit
vector $\bc \in \R_j^2\cap \R_k^2$.

 Expanding $e^{\omega \bc}=e^{-\phi \ba}e^{\rho \bb}$, we get
  \[ \cosh \omega (1 + \bv_w ) =e^{-\phi \ba}e^{\bb}=\cosh \phi \cosh \rho(1 -\bu_v )(1 +\bu_w) \]
\[ =\cosh \phi \cosh \rho [(1-\bu_v \cdot \bu_w)+(\bu_w - \bu_v  - \bu_v \w \bu_w)]. \]
It follows that 
   \beq \cosh \omega =(\cosh \phi \cosh \rho)(1-\bu_v \cdot \bu_w), \label{coshomega} \eeq
 and 
  \beq  \bv_w =\frac{\cosh \phi \cosh \rho}{\cosh \omega } (\bu_w  - \bu_v  
- \bu_v \w \bu_w )=\frac{\bu_w  - \bu_v  
- \bu_v \w \bu_w }{1-\bu_v \cdot \bu_w} .  \label{sinhomega}  \eeq  
     
   %  [\cosh^2 \phi \sinh^2 \rho + \sinh^2 \phi \cosh^2 \rho - 2 \ba \cdot \bb \cosh \phi 
    %       \sinh \rho \cosh \rho \sinh \phi + (\ba \w \bb)^2 \sinh^2 \phi \sinh^2 \rho]^\frac{1}{2} 
 %\bb \cosh \phi \sinh \rho - \ba \sinh \phi \cosh \rho - \ba \w \bb \sinh \phi \sinh \rho|
 
   We have found a relative unit vector $\bc \in \R_j^2\cap \R_k^2$,
$ \bc:=\frac{\bv_w}{\tanh \omega}$ and a hyperbolic angle $\omega \ge 0$ with
the property that
  \[ k = j e^{\omega \bc} = e^{-\frac{1}{2}\omega \bc}j e^{\frac{1}{2}\omega \bc}  . \]
The relative bivector $k$ has velocity $\bv_w=\bc \tanh \omega$ with respect to
the $j$. However, the relative unit vector $\bc \notin \R_i^2$. This means that
the relative vector $\bc$ defining the direction of the velocity of the relative bivector $k$ with respect
to $j$ is not {\it commensurable} with the vectors in $\R_i^2$. 

The question arises whether or not there exist
a unit vector $\bd \in \R_i^2$ with the property that
   \beq  k=e^{-\frac{1}{2}\Omega \bd}j e^{\frac{1}{2}\Omega \bd}. \label{irelvel} \eeq
Substituting $j=ie^{\phi \ba}$ and $k=i e^{\rho \bb}$ into this last equation gives
  \[ i e^{\rho \bb}=e^{-\frac{1}{2}\Omega \bd}ie^{\phi \ba} e^{\frac{1}{2}\Omega \bd}, \]
 which is equivalent to the equation
  \beq   e^{\rho \bb}=e^{\frac{1}{2}\Omega \bd}e^{\phi \ba} e^{\frac{1}{2}\Omega \bd}. \label{equirelvel}\eeq
The transformation $L_p:\G_2 ->\G_2$ defined by  
  \beq L_p(\bx)= e^{\frac{1}{2}\Omega \bd}\bx e^{\frac{1}{2}\Omega \bd} \label{pboost} \eeq
is called the {\it passive Lorentz boost} relating $\R_j^2$ to $\R_k^2$ with respect to $\R_i^2$.

The equation (\ref{equirelvel}) can either be solved for $ e^{\rho \bb}$ given $ e^{\Omega \bd}$ and
$ e^{\phi \ba}$, or for $ e^{ \Omega \bd}$ given $ e^{\phi \ba}$ and
$ e^{\rho \bb}$. Defining the velocities $\bu_v = \ba \tanh \phi, \bu_w= \bb \tanh \rho$, and
$\bu_{vw}=\bd \tanh \Omega$, we first solve for $ e^{\rho \bb}$ given $ e^{\Omega \bd}$ and
$ e^{\phi \ba}$. In terms of these velocities, equation (\ref{equirelvel}) takes the form
  \[ \cosh \rho \ (1+\bu_w) = \cosh \phi \ e^{\frac{1}{2}\Omega \bd}(1+\bu_v) e^{\frac{1}{2}\Omega \bd} 
   = \cosh \phi \ (e^{\Omega \bd} +  e^{\frac{1}{2}\Omega \bd}\bu_v e^{\frac{1}{2}\Omega \bd})  \]
   \[ = \cosh \phi \cosh \Omega \ [(1+\bu_{vw})(1+ \bu_v^\parallel)]+ \cosh \phi \ \bu_v^\perp \]
 where $\bu_v^\parallel = (\bu_v \cdot \bd)\bd$ and  $\bu_v^\perp = (\bu_v \w \bd)\bd$ . 
 Equating scalar and vector parts gives
    \beq \cosh \rho = \cosh \phi \cosh \Omega \ (1+\bu_v \cdot \bu_{vw}),  \label{HesSca}  \eeq
and 
   \beq   \bu_w = \frac{\bu_v+\bu_{vw}+(\frac{1}{\cosh \Omega} -1) (\bu_v \w \bd)\bd}{1+\bu_v\cdot \bu_{vw}} .
                                               \label{HesAdd}   \eeq
The equation (\ref{HesAdd}) is the (passive) composition formula for the addition of velocities of special relativity 
in the system $u$, \cite[p.588]{H99} and \cite[p.133]{LP01}.                                               
                                               
    To solve (\ref{equirelvel}) for $ e^{ \Omega \bd}$ given $ e^{\phi \ba}$ and
$ e^{\rho \bb}$, we first solve for the unit vector $\bd \in \R^2$ by taking the anti-symmetric
product of both sides of (\ref{equirelvel}) with $\bd$ to get the relationship
  \[  \bd \otimes \bb\ \sinh \rho = e^{\frac{1}{2}\Omega \bd}\bd \otimes \ba\ \sinh \phi \ e^{\frac{1}{2}\Omega \bd}=
                              \ba \sinh \phi             , \]
or equivalently,
  \[ \bd \w (\bb \sinh \rho - \ba \sinh \phi)= 0 . \]
In terms of the velocity vectors $\bu_v$ and $\bu_w$, we can define the unit vector $\bd$ by 
  \beq \bd = \frac{ \bu_w \cosh \rho - \bu_v \cosh \phi}{\sqrt{\bu_v^2 \cosh^2 \phi - 2 \bu_v \cdot \bu_w
 \cosh \phi \cosh \rho+\bu_w^2  \cosh^2 \rho}} \label{ddir} \eeq  
                      
   Taking the symmetric product of both sides of (\ref{equirelvel}) with $\bd$ gives
  \[  [ \bd \circ e^{\phi \ba}]e^{\Omega \bd}=\bd \circ e^{\rho \bb} , \]
or 
  \[ (\bd \cosh \phi + \ba \cdot \bd \sinh \phi)e^{\Omega \bd}
= \bd \cosh \rho + \bb\cdot \bd \sinh \rho . \]
Solving this last equation for $e^{\Omega \bd}$ gives
  \beq e^{\Omega \bd}= \frac{(\bd \cosh \rho + \bb\cdot \bd \sinh \rho)(\bd \cosh \phi - \ba\cdot \bd \sinh \phi)}
        {\cosh^2 \phi -(\ba \cdot \bd)^2\sinh^2 \phi},   \label{eOmega}\eeq 
or in terms of the velocity vectors,
        \[ \cosh \Omega (1 + \bu_{vw})=\frac{\cosh \rho}{\cosh \phi}\Big( \frac{(1+\bu_w \cdot \bd \ \bd)
                       (1-\bu_v \cdot \bd \ \bd)}{1-(\bu_v \cdot \bd)^2}\Big) \]
       \[ = \frac{\cosh \rho}{\cosh \phi}\Big(\frac{1 - (\bu_v \cdot \bd)(\bu_w \cdot \bd)+
                       (\bu_w - \bu_v)\cdot \bd \ \bd}{1-(\bu_v \cdot \bd)^2}\Big) . \]                
 Taking scalar and vector parts of this last equation gives
    \beq  \cosh \Omega= \frac{\cosh \rho}{\cosh \phi}\Big(\frac{1 - (\bu_v \cdot \bd)(\bu_w \cdot \bd)}
    {1-(\bu_v \cdot \bd)^2}\Big)  \label{coshOmega}                \eeq 
  and
   \beq   \bu_{vw} =  \frac{    (\bu_w - \bu_v)\cdot \bd \ \bd}{1-(\bu_v \cdot \bd)(\bu_w\cdot \bd)} 
     \label{sinhOmega}.   \eeq 
    
     We say that $\bu_{vw}$ is the relative velocity of the passive boost (\ref{pboost}) of $j$ into
$k$ relative to $i$. The passive boost is at the foundation of the {\it Algebra of Physical Space} formulation
of special relativity \cite{BS}, and a coordinate form of this passive approach was used by Einstein in his 
famous 1905 paper \cite{E1905}. Whereas Hestenes in \cite{H03} employs the active Lorentz boost, in \cite{H99}
he uses the passive form of the Lorentz boost.

The distinction between active and passive boosts continues to be the source of much confusion in the literature
\cite{O05}. Whereas an active boost (\ref{boost}) mixes vectors and bivectors of $\G_2$, the passive boost
defined by (\ref{pboost}) mixes the vectors and scalars of $\G_2$ in the geometric algebra $\G_2$ of $i$.   
In the next section, we shall find an interesting geometric interpretation of this
result in a closely related higher dimensional space.

\section*{Splitting the plane}

  Geometric insight into the previous calculations can be obtained by {\it splitting} or {\it factoring} the geometric
algebra $\G_2$ into a larger geometric algebra $\G_{1,2}$. The most mundane way of accomplishing this is to
factor the standard orthonormal basis vectors of (\ref{stdbasis}) into an orthonormal bivectors of a larger
geometric algebra $\G_{1,2}$. We write
  \[ \be_1 = \gamma_0 \gamma_1, \ \ {\rm and} \ \ \be_2=\gamma_0 \gamma_2 , \]
and assume the rules $\gamma_0^2 = 1 = -\gamma_1^2 = -\gamma_2^2$, and $\gamma_\mu \gamma_\eta=-\gamma_\eta \gamma_\mu$
for all $\mu, \eta = 0,1,2$ and $\mu \ne \eta$. The standard orthonormal basis of $\G_{1,2}$ consists of the eight
elements
  \[ \{1, \gamma_0, \gamma_1, \gamma_2, \gamma_{01}, \gamma_{02},\gamma_{21}, \gamma_{012} \} . \]
With this splitting, the standard basis elements (\ref{stdbasis}) of $\G_2$ are identified with 
elements of the {\it even subalgebra}
  \beq \G_{2,1}^+ := span \{1, \be_1=\gamma_{01},\be_2 =\gamma_{02},\be_{12}=\gamma_{21} \}  \label{evensubalg} \eeq
of $\G_{1,2}$. We denote the oriented unit {\it pseudoscalar} element by $s=\gamma_{012}$. Note
that $s\in Z(\G_{1,2})$, the {\it center} of the algebra $\G_{1,2}$.

  Consider now the mapping 
   \beq \psi : {\cal H}^+ \longrightarrow \{ r \in \G_{1,2}^1| \ \ r^2=1 \} 
                                                                     \label{sduality} \eeq
defined by $ r=\psi(h)=s h$  for all $h\in {\cal H}^+$.   The mapping $\psi$ sets up a $1-1$ correspondence between the 
positively oriented unit bivectors $h\in {\cal H}^+$ and
unit timelike vectors $r \in \G_{1,2}^1$ which are {\it dual} under multiplication by the pseudosclar $s$.
Suppose now that $\psi (i)=u$, $\psi(j)=v$ and $\psi(k)=k$. Then it immediately follows by duality that if
$j=ie^{\phi \ba}$, $k=ie^{\rho \bb}$ and $k=je^{\omega \bc}$, then $v = ue^{\phi \ba}$, $w = ue^{\phi \ba}$ and
$w =v e^{\omega \bc}$, respectively. It is because of this $1-1$ correspondence that we have included the
labels $u$, $v$ and $w$ as another way of identifying the oriented planes of the bivectors $i$, $j$ and $k$ in Figure 3.

  Just as vectors in $\bx, \by \in \G_2^1$ are identified with points $\bx, \by \in \R^2$, 
{\it Minkowski vectors} $x,y \in \G_{1,2}^1$ are identified with points $x,y \in \R^{1,2}$ 
the $3$-dimensional pseudoeuclidean
space of the {\it Minkowski spacetime plane}. A Minkowski vector $x \in \R^{1,2}$ is said to be {\it timelike} if
$x^2 >0$, {\it spacelike} if $x^2 <0$, and {\it lightlike} if
$x\ne 0$ but $x^2=0$. For two Minkowski vectors $x, y \in G_{1,2}^1$, we decompose the geometric product $xy$
into {\it symmetric} and {\it anti-symmetric} parts
  \[  xy = \frac{1}{2}(xy+yx)+\frac{1}{2}(xy-yx)=x\cdot y+x\w y,\]
where $x\cdot y:= \frac{1}{2}(xy+yx)$ is called the {\it Minkowski inner product} and 
$x\w y:=\frac{1}{2}(xy-yx)$ is called the {\it Minkowski outer product} to distinquish these
products from the corresponding inner and outer products defined in $\G_2$. 

In \cite{H74} and \cite{H03}, David Hestenes gives an active reformulation of Einstein's special relativity in the
spacetime algebra $\G_{1,3}$. In \cite{S2,S3}, I show that an equivalent active reformulation is possible
in the geometric algebra $\G_3$ of the Euclidean space $\R^3$. In \cite{BS} and \cite{Alex02}  
the relationship between active and passive formulations is considered.

For the two unit timelike vectors $u,v \in \G_{1,2}$, we have
  \beq  uv= u\cdot v + u\w v = e^{\phi \ba} = \cosh \phi + \ba \sinh \phi . \label{hypgeoproduct} \eeq  
It follows that $u\cdot v = \cosh \phi$ and $u\w v= \ba \sinh \phi$, which are the 
hyperbolic counterparts to the geometric product of unit vectors $\ba, \bb \in \R^2$ given in (\ref{geoproduct}). 
%Note that $u\w v=\ba \sinh \phi$ shows that that the bivector $u\w v \in \G_{1,2}^2$
%is the Euclidean vector $\ba \sinh \phi \in \G_2$. 
The Minkowski bivector
  \[ \bu_v= \frac{u\w v}{u\cdot v}=\ba \tanh \phi = -\bv_u \]
 is the {\it relative velocity} of the timelike vector unit vector $v$ in the frame of $u$. 
 
  Suppose now that for $i,j,k \in {\cal H}^+$, $\psi(i)=u, \psi(j)=v$ and $\psi(k)=w$, so that
$uv=e^{\phi \ba }, uw = e^{\rho \bb }$ and $vw=e^{ \omega \bc}$, respectively. Let us recalculate
$vw=e^{\omega \bc }$ in the {\it spacetime algebra} $\G_{1,2}$:
   \[  vw = vuuw =(vu)(uw)=(v\cdot u-u\w v)(u \cdot w +u\w w) \]
   \[= (v\cdot u)(w\cdot u)(1-\bu_v)(1+\bu_w) .\]
Separating into scalar and vector parts in $\G_2$, we get 
   \beq v\cdot w = (v\cdot u)(w\cdot u)(1-\bu_v\cdot \bu_w )  \label{coshomeganew} \eeq
and  
   \beq (v\cdot w)\bv_w=(v\cdot u)(w\cdot u)[\bu_w - \bu_v - \bu_v\w \bu_w ],  \label{sinhomeganew} \eeq
 identical to what we calculated in (\ref{coshomega}) and (\ref{sinhomega}), respectively. 

More eloquently, using (\ref{coshomeganew}), 
we can express (\ref{sinhomeganew})  in terms of quantities totally in the algebra $\G_2$,   
   \beq \bv_w=\frac{\bu_w - \bu_v  -\bu_v\w \bu_w }{1-\bu_v \cdot \bu_w} = \bc \tanh \omega =-\bw_v.  \label{eloquent} \eeq
We see that the relative velocity $\bv_w$, up to a scale factor, 
 is the {\it difference} of the velocities $\bu_w$ and  $\bu_v$ and the bivector $\bu_v\w \bu_w $ in the 
 system $u$. Setting $w=v$ in (\ref{coshomeganew}) and solving for $v \cdot u$ in terms of $\bu_v^2$ gives
   \beq  u\cdot v = \frac{1}{\sqrt{1-\bu_v^2}},  \label{vdotu} \eeq
a famous expression in Einstein's theory of special relativity, \cite{H03}.
 
    Let us now carry out the calculation for (\ref{coshOmega}) and the relative velocity (\ref{sinhOmega}) of the system
 $w$ with respect to $v$ as measured in the frame of $u$. We begin by defining the bivector
$D=(w-v)\w u$ and noting that $w\w D = v \w w \w u = v\w D$. Now note that
  \[ w= w D D^{-1}=(w\cdot D)D^{-1} + (w\w D)D^{-1}=w_\parallel + w_\perp  \]
where $ w_\parallel = (w\cdot D)D^{-1}$ is the component of $w$ {\it parallel} to $D$ and 
$ w_\perp = (w\w D)D^{-1}$ is the component of $w$ {\it perpendicular} to $D$. Next, we
calculate
  \beq \hat w_\parallel \hat v_\parallel= - \frac{(w\cdot D)(v\cdot D)}{|w\cdot D||v\cdot D|}
     =\frac{(w\cdot D)(v\cdot D)}{(v\cdot D)^2} 
     =(w\cdot D)(v \cdot D)^{-1}, \label{ucal}  \eeq
since 
   \[ (w\cdot D)^2=(v\cdot D)^2 =[(w\cdot v -1)u-(v\cdot u)(w-v)]^2  \]
   \[ =   (w\cdot v -1)[(w\cdot v -1)-2(v\cdot u)(w\cdot u)]<0.  \]
  
   We can directly relate (\ref{ucal}) to  (\ref{coshOmega}) and (\ref{sinhOmega}),
     \[ \hat w_\parallel \hat v_\parallel= \frac{(w\cdot D)uu(v\cdot D)}{(v\cdot D)^2}=
      \frac{[(w\cdot D)\cdot u+(w\cdot D)\w u ][(v\cdot D)\cdot u+(v\cdot D)\w u ]}{(v\cdot D)^2} \]
    \[ = \frac{[-(w\w u)\cdot D- (w\cdot u)D ][-(v\w u)\cdot D+ (v\cdot u)D ]}{(v\cdot D)^2} \]   
\[ =-(w\cdot u)(v\cdot u) \frac{[\bu_w \cdot \bd + \bd ][-\bu_v\cdot \bd+ \bd ]}{(v\cdot \bd)^2} \]
  \[ = -(w\cdot u)(v\cdot u) \frac{1-(\bu_v\cdot \bd )( \bu_w \cdot \bd) 
               +(\bu_w-\bu_v)\cdot \bd \ \bd}{(v\cdot \bd)^2}  \]
where we have used the fact that $\bd=D/|D|$, see (\ref{ddir}). In the special case when $v=w$, the above equation
reduces to $(v \cdot \bd)^2=-(u \cdot v)^2[1-(\bu_v\cdot \bd)^2]$. Using this result in the previous
calculation, we get the desired result that
 \[ \hat w_\parallel \hat v_\parallel= \frac{(w\cdot u)[1-(\bu_v\cdot \bd )( \bu_w \cdot \bd) 
               +(\bu_w-\bu_v)\cdot \bd \ \bd]}{(u \cdot v)[1-(\bu_v\cdot \bd)^2]} , \]
 the same expression we derived after equation (\ref{eOmega}).
  
 \begin{figure}
\begin{center}
\includegraphics[scale=0.9]{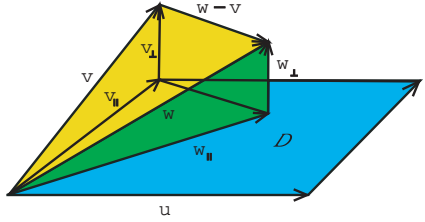}
\caption{Passive boost in the spacetime plane of $D$.}
\end{center}
\end{figure}   

    Defining the active boost 
$L_u(x)=(\hat w_\parallel \hat v_\parallel)^{\frac{1}{2}}x(\hat v_\parallel \hat w_\parallel )^{\frac{1}{2}}$,
we can easily check that it has the desired property that
  \[ L_u(v)=L_u(v_\parallel+v_\perp) =\hat w_\parallel \hat v_\parallel v_\parallel +v_\perp = w_\parallel+w_\perp = w .\] 
    Thus, the {\it active} boost taking the unit timelike vector $\hat v_{\parallel}$ into
    the unit timelike vector $\hat w_{\parallel}$ is equivalent to the passive boost (\ref{pboost}) in the
    plane of the spacetime bivector $D$. See Figure 4.

The above calculations show that each different system $u$ measures {\it passive relative velocities} between
the systems $v$ and $w$ differently by a boost in the plane of the Minkowski bivector $D=(w-v)\w u$, whereas
there is a {\it unique} active boost (\ref{boost}) that takes the system $v$ into $w$ in the plane of $v\w w$.
The concept of a passive and active boost become equivalent when $u\w v\w w = 0$, the case when $\bb = \pm \ba$.

\section*{Appendix: Matrix Representation}

   The algebraic rules satisfied by elements of $\G_2$ are completely compatible with the rules of matrix
algebra and provide a {\it geometric basis} for $2 \times 2$ matrices. 
By the {\it spectral basis} of $\G_2$ we mean 
  \beq \pmatrix{1 \cr \be_1 } u_+ \pmatrix{1 & \be_1} = \pmatrix{ u_+ & \be_1 u_- \cr \be_1 u_+ & u_-}, \label{spectralbasis} \eeq 
where $u_\pm = \frac{1}{2}(1\pm \be_2)$ are mutually annihiliating idempotents, \cite{GS2004}. 
 
Noting that
  \[ \pmatrix{1 & \be_1 } u_+ \pmatrix{1 \cr \be_1} = u_+ + \be_1 u_+ \be_1 = u_+ + u_- =1 ,\]
for $g=x+iy+v_1 \be_1+v_2 \be_2 \in \G_2$, we find that 
   \[g=\pmatrix{1 & \be_1 } u_+ \pmatrix{1 \cr \be_1}g \pmatrix{1 & \be_1 } u_+ \pmatrix{1 \cr \be_1}  \]
     \[=\pmatrix{1 & \be_1 } u_+ \pmatrix{g & g \be_1 \cr \be_1 g & \be_1 g \be_1} u_+ \pmatrix{1 \cr \be_1 } \]
   \[=\pmatrix{1 & \be_1 } u_+ \pmatrix{x+v_2 & v_1-y \cr v_1+y & x-v_2}\pmatrix{1 \cr \be_1 } .\]
The real matrix $[g]:= \pmatrix{x+v_2 & v_1-y \cr v_1+y & x-v_2}$ is called the
{\it matrix} of $g$ with respect to the spectral basis (\ref{spectralbasis}).
   
   By the {\it inner automorphism} or {\it $\be_1$-conjugate} 
   $g^{\be_1}$ of $g\in \G_3$ with respect to the unit vector $\be_1$,
  we mean 
   \beq g^{\be_1} :=\be_1 g \be_1 .   \label{econjugate} \eeq
   We can now explicitly solve for the matrix $[g]$ of $g$. 
     \[  \pmatrix{1 \cr \be_1 } g \pmatrix{1 & \be_1} =\pmatrix{1 & \be_1 \cr \be_1 &1 } u_+[g] \pmatrix{1 & \be_1 \cr \be_1 & 1} \]
 or  
   \[  u_+ \pmatrix{1 \cr \be_1 } g \pmatrix{1 & \be_1}u_+ =u_+ \pmatrix{1 & \be_1 \cr \be_1 &1 } u_+[g]
    \pmatrix{1 & \be_1 \cr \be_1 & 1}u_+ = u_+[g], \]
 and taking the $\be_1$-conjugate of this equation gives   
  \[  u_- \pmatrix{1 \cr \be_1 }  g^{\be_1} \pmatrix{1 & \be_1}u_- =u_-[g]. \]
Adding the last two expressions gives the desired result that 
    \[ [g]= u_+ \pmatrix{g & g \be_1 \cr \be_1 g & \be_1 g \be_1} u_+ + u_-
     {\pmatrix{\be_1 g \be_1 & \be_1 g  \cr g \be_1 &  g }} u_- .\]  
    
   Of course, the geometric numbers of the spacetime algebra $\G_{1,2}$ also have a matrix representation.
Since the unit pseudoscalar element $s=\gamma_{012}\in \G_{1,2}$ is in the center of the algebra,
$s\in Z(\G_{1,2})$ and $s^2=-1$, it follows that a general element $f\in \G_{1,2}$ can be expressed
as the {\it complexification} of the algebra $\G_3$. Thus, we write $f=g+sh$ for $g,h \in \G_2$.
Then for $g=x_1+iy_1+a_1 \be_1+a_2 \be_2$ and $h=x_2+iy_2+b_1 \be_1+b_2 \be_2$, we have 
  \[ [f] = [g+sh] = [g]+s[h]=  \pmatrix{x_1+a_2 & a_1-y_1 \cr a_1+y_1 & x_1-a_2}+
   s \pmatrix{x_2+b_2 & b_1-y_2 \cr b_1+y_2 & x_2-b_2}. \] 
   
   The larger geometric algebra $\G_{1,2}$ has three involutions which
are related to complex conjugation. The {\it main involution} is obtained by changing the sign of
all vectors in $\G_{1,2}^1$. For $f=g+sh$, $f^*:=g-sh$. The main involution thus behaves
as the complex conjugation of the pseudoscalar $s$.
 {\it Reversion} is obtained by reversing the order of the products of vectors in $\G_{1,2}^1$.
For $f=g+sh$ given above, $f^\dagger = g^\dagger - s h^\dagger$
where $g^\dagger = x_1-iy_1-a_1 \be_1-a_2 \be_2$ and $h^\dagger = x_2-iy_2-b_1 \be_1-b_2 \be_2$ 
The third involution, called {\it Clifford conjugation} is obtained by combining the above two operations,
   \beq \wt f :=(f^*)^{\dagger}= {g^*}^\dagger -s {g^*}^\dagger  \label{conjugation}, \eeq
where ${g^*}^\dagger = x_1-iy_1-a_1 \be_1-a_2 \be_2$ and ${h^*}^\dagger = x_2-iy_2-b_1 \be_1-b_2 \be_2$ .
  
   Finally, we note that the geometric algebra $\G_{1,2} $ is algebraically closed with $j=\gamma_{012}\in Z(\G_{1,2})$.
This means that in dealing with the characteristic and minimal polynomials of the matrices which
represent the elements of $\G_{1,2}$, we can always interpret complex zeros of these polynomials to be
in the spacetime algebra $\G_{1,2}$.

  \section*{Acknowledgements} 
     The author thanks David Hestenes for teaching him special relativity in the beautiful language of spacetime
algebra many years ago at Ariziona State University. The author is grateful to Dr. Andres Ramos, and
Dr. Guillermo Romero of the Universidad
de Las Americas for support for this research. He also thanks William Baylis and Zbygniew Oziewicz for
discussions about relative velocity. The author is is a member of SNI 14587. 
(URL: http://www.garretstar.com)

\end{document}